\documentclass[11pt]{article}
\usepackage{latexsym,html,graphics,pictex,epsfig}  
\newcommand\be{\begin{equation}}
\newcommand\ee{\end{equation}}
\newcommand{\fa}{\begin{eqnarray}}
\newcommand{\ffa}{\end{eqnarray}}

\def\Comp{{\mathchoice
{\setbox0=\hbox{$\displaystyle\rm C$}\hbox{\hbox to0pt
{\kern0.4\wd0\vrule height0.9\ht0\hss}\box0}}
{\setbox0=\hbox{$\textstyle\rm C$}\hbox{\hbox to0pt
{\kern0.4\wd0\vrule height0.9\ht0\hss}\box0}}
{\setbox0=\hbox{$\scriptstyle\rm C$}\hbox{\hbox to0pt
{\kern0.4\wd0\vrule height0.9\ht0\hss}\box0}}
{\setbox0=\hbox{$\scriptscriptstyle\rm C$}\hbox{\hbox to0pt
{\kern0.4\wd0\vrule height0.9\ht0\hss}\box0}}}}
\def\C{\Comp}

\def\Et{\tilde{E}}
\def\htt{\tilde{\tilde{h}}}
\def\ifi{\infty}
\def\sbs{\subset}

\def\Si{\Sigma}
\def\G{\Gamma}
\def\D{\Delta}
\def\O{\omega}
\def\p{\psi}
\def\b{\beta}
\def\r{\rho}
\def\e{\epsilon}
\def\g{\gamma}
\def\a{\alpha}
\def\d{\delta}
\def\l{\lambda}

\begin{document}

\title{Classical Geometry from a Physical State in Canonical Quantum
Gravity}
\author{Yongge Ma$^1$\thanks{E-mail: ygma@fis.uncor.edu}
and Yi Ling${}^{2}$\thanks{E-mail: ling@phys.psu.edu}\\
{ }\\
\centerline{\it ${}^1$FaMAF, Universidad Nacional de Cordoba,}\\
\centerline{\it 5000, Cordoba, Argentina}\\
{ }\\ \centerline{\it
${}^2$Center for Gravitational Physics and Geometry}\\
\centerline{\it Department of Physics}\\ \centerline {\it The
Pennsylvania State University}\\ \centerline{\it University Park,
PA, USA 16802}}

\maketitle

\begin{abstract}
\baselineskip=20pt
We construct a weave state which approximates a degenerate 3-metric of rank 2 
at large scales. It turns out that a non-degenerate metric region can be 
evolved from this degenerate metric by the classical Ashtekar equations, 
hence the degeneracy of 3-metrics is not preserved by the evolution of 
Ashtekar's equations. As the s-knot state corresponding to this weave is 
shown to solve all the quantum constraints in loop quantum gravity, a 
physical state in canonical quantum gravity is related to the familiar 
classical geometry.

\end{abstract}

\newpage

\section{Introduction}
\label{sec:1} \baselineskip=20pt

Since the new Hamiltonian formulation of gravity was proposed by Ashtekar in
1986 \cite{As86}, considerable progress has been made in non-perturbative
canonical quantum gravity, namely, loop quantum gravity \cite{Ro98}. One of the
surprising results of the theory is that some solutions of all the quantum
constraints in canonical gravity have been found \cite{RS90,Ez96}. The 
kinematic of the theory is now rigorously defined \cite{AL95,Ba96}. However,
to accept the theory as a conceivable candidate for describing quantum
space-time, we need to prove that its classical limit is general relativity(GR)
or at least overlaps GR in the regime where GR is well tested.

A weave state was first introduced in Ref.\cite{ARS92} to approximate the flat
 3-geometry. Being one of the solutions in Ref.\cite{RS90}, it is regarded as 
a physical state of loop quantum gravity. However, this state is an 
eigenstate of the volume operator\cite{RS95,AL98} with vanishing eigenvalue. 
Moreover, as argued in Refs.\cite{Ma94,Ez96}, the classical correspondences of 
the solutions in Ref.\cite{RS90} and their generalisation\cite{Ez96} should 
all be degenerate metrics which are not admitted in the traditional GR. 
While, as far as we know, the other weave states appearing so far are all 
kinematical states at the unconstrained level \cite{Ze92,Bo94,GR96,AG99}.

As there is growing evidence from various descriptions of quantum gravity 
that degenerate metrics should have an important role \cite{RS90,AW88,Ho91}, 
researche has been promoted to study the degenerate metric in the classical 
Ashtekar theory as it is admitted by the formalism, including the 
dynamic characters of degenerate triads\cite{Ja96,LW979,ML98b} and 
degenerate phase boundaries\cite{BJ97,MLK99,ML99}. By breaking the causality, 
a solution to classical Ashtekar's equations was constructed in 
Ref.\cite{ML98a}, where a degenerate space-time region could be evolved from 
non-degenerate initial data. However, no example has been raised so far where 
a non-degenerate metric is generated by the time evolution of degenerate 
initial data, although it is not impossible in principle.

The present paper involves the above two topics. We will show that, in loop 
quantum gravity a quantum state based on the s-knot class of an infinite 
number of open curves can solve all quantum constraints and approximate 
a degenerate 3-geometry, from which a non-degenerate metric region can be 
evolved by the classical Ashtekar equations. Thus, a physical state of 
canonical quantum gravity is related to the familiar classical geometry. In 
Section 2, we construct a ``quasi-coherent'' state and show how it can weave 
a rank 2 degenerate metric on $R^3$ at large scales. In section 3, we show an 
example in complex Ashtekar's formalism where a non-degenerate metric is 
generated by the time evolution of the degenerate 3-metric. The physical state 
related to the weave is induced after a few comments and discussions in 
Section 4.

\section{Weaving a degenerate metric}
\label{sec:2}

\subsection{Preliminaries}
\label{sec:2.1}

Canonical gravity in the real Ashtekar formalism is defined over an oriented 
3-manifold $\Si$ \cite{Ba95}. The basic variables are real $SU(2)$ connections 
$A_a^i$ and densitized triads $\Et^b_j$ of weight 1. We use $a,b,\dots =1,2,3$ 
for spatial indices and $i,j,\dots =1,2,3$ for internal $SU(2)$ indices. A 
tilde over(under) a letter denotes a density weight 1(-1). 

The Ashtekar theory admits a generalisation of GR to involve degenerate 
metrics, since the inverse of the triads is not necessary for the whole 
formalism. The triad is related to the 3-metric by
\begin{equation}
\label{1}
\htt^{ab}=\Et^a_i\Et^{bi}.
\end{equation}
In the case where $h_{ab}$ is non-degenerate,
\be \label{2}
\left( \htt^{ab}\right)=h(h^{ab})=({\cal A}^{ab}),
\ee
where $h:=det(h_{ab})$, and the elements of $({\cal A}^{ab})$ are the
cofactors of $(h_{ab})$. Note that Eq.(\ref{2}) can be naturally generalised
to the case where $h_{ab}$ becomes degenerate by neglecting the middle
procedure.

Based on the Ashtekar variables, canonical quantum gravity can also be 
represented by loop variables, and there exist well-defined non-local 
operators carrying geometric informations \cite{RS90,ARS92}. In the following 
we will consider the operator $\hat{Q}[\O]$ associated with one-forms $\O_a$ 
on $\Si$ \cite{ARS92,GR96}, rather than the operators of area and volume. One 
can see, $\hat{Q}[\O]$ is more suitable to respect the feature of a 
degenerate metric, as it is related to the classical quantity
\be \label{4}
Q[\O]= \int d^3x \sqrt{\Et^a_i\O_a\Et^{bi}\O_b}= \int d^3x
\sqrt{\htt^{ab}\O_a\O_b},
\ee
where $\O_a$ is any smooth 1-form which makes the integral meaningful and the
integral is well defined since the integrand is a density of weight 1. Using 
loop variables \cite{RS90,PR96}, the same quantity could be expressed as 
\be \label{5}
Q[\O]= \lim_{\e\rightarrow 0} \int d^3x \left[\int d^3y \int d^3z
f_\e(x,y)f_\e(x,z){1 \over 2}{\cal T}^{ab}[\a_{yz}](y,z)\O_a(y)\O_b(z)\right]^
{1\over 2},
\ee
where $f_\e(x,y)$ tends to $\d(x,y)$ as $\e$ tends to zero, $\a_{yz}$ is an
arbitrarily defined smooth loop in $\Si$ that passes through points $y$ and $z$
such that it goes to a point as $y\rightarrow z$, and the loop variable
\be\label{6}
{\cal T}^{ab}[\a](y,z)=-Tr[\r_1(H_{\a}(y,z)\Et^b(z)H_{\a}(z,y)\Et^a(y))],
\ee
here $H_{\a}(y,z):={\cal P}exp\left[-\int_z^y ds \dot{\a}^aA_a(\a(s))
\right]$ is the
holonomy or parallel propagator of the connection along $\a$, and the
2-dimensional representation, $\r_1$, of $SU(2)$ is used to evaluate traces. 
Eq.(\ref{5}) is valid for the quantum version: One can get the well-defined 
quantum operator $\hat{Q}[\O]$ simply by replacing ${\cal T}^{ab}$ by the loop 
operator $\hat{{\cal T}}^{ab}$ \cite{ARS92,PR96}. The action of this operator 
on a coloured loop state gives\cite{GR96}
\be\label{7} 
\hat{Q}[\O] \Psi_P[\g]=16\pi
l_{Pl}^2\sqrt{{P \over 2}({P \over 2}+1)}\int_\g ds
|\dot{\g}^a\O_a(\g(s))|\Psi_P[\g], 
\ee
where $P$ is the positive integer associated with the loop $\g$, $l_{Pl}$ 
denotes the Planck length, and $\Psi_P[\g]:=Tr[\r_P(H[\g])]$, here $\r_P$ 
denotes the ($P+1$)-dimensional representation of $SU(2)$.

We now briefly introduce the Hilbert space of loop quantum gravity 
\cite{AL95,Ba96}. Given any graph, $\G_n=\{\g_1, \dots, \g_n\}$, embedded in 
$\Si$ and a function $f_n: [SU(2)]^n\rightarrow \C$, the cylindrical function 
is defined as:
\be\label{8}
\Psi_{\G_n,f_n}(A):=f_n(H[\g_1], \dots, H[\g_n]).
\ee
Since any two cylindrical functions based on different graphs can always be
viewed as being defined on the same graph which is just constructed as the
union of the original ones, it is straightforward to define a scalar product
for them by:
\be\label{9}
\langle \Psi_{\G_n,f_n}|\Psi_{\G_n,g_n}\rangle := \int_{[SU(2)]^n} dH_1\dots 
dH_n \overline{f_n(H_1,\dots, H_n)}g_n(H_1, \dots, H_n),
\ee
where $dH\dots dH_n$ is the Haar measure of $[SU(2)]^n$ which is naturally
induced by that of $SU(2)$. The Hilbert space, $\cal H$, is obtained by
completing the space of all finite linear combinations of cylindrical functions
in the norm induced by the quadratic form (\ref{9}) on a cylindrical function.

The operators of area and volume have been shown to be self-adjoint on 
$\cal H$ \cite{AL97,AL98}. Since $\hat{Q}[\O]$ is closely related to 
the area operator \cite{GR96}, it is reasonable to conceive it is also 
self-adjoint. This conceit can be rigorously proved \cite{ML00}.

\subsection{The weave}
\label{sec:2.2}

The geometry which we want to approximate is a degenerate "flat" 3-metric,
$h_{ab}$, of rank 2 on $R^3$. The metric is "flat" in the sense that there
exists a foliation $R^3=R^2\times R$ such that the induced 2-metric, $q_{ab}$,
of $h_{ab}$ on $R^2$ is the flat Euclidean metric.

Let $\{ X, Y, Z \}$ be the Cartesian coordinates on $R^3$ compatible with the
decomposition $R^3=R^2\times R$ and $({\partial\over\partial Z})^a$ be the
degenerate vector field of $h_{ab}$. Thus the line element of $h_{ab}$
reads
\be \label{3}
ds^2=dX^2+dY^2.
\ee
Hence, from Eq.(\ref{2}) the only non-zero component of $\htt^{ab}$ is
$\htt^{ZZ}=1$.

The weave states which approximate classical 3-metrics were first constructed
as the eigenstates of geometrical operators such as $\hat{Q}[\O]$ and the
operators of area and volume \cite{ARS92,Ze92,GR96}. The corresponding
eigenvalues are required to agree with the classical values of the geometrical
quantities at large scales. The updated successful construction of $\cal H$
promotes us now to approximate a classical geometry by the expectation values
of the geometrical operators.

For the operator $\hat{Q}[\O]$, one can define the following: A quantum state 
$\Psi$ is said to approximate a classical metric on $\Si$ at scales
larger than a macroscopic length scale $L$ accessible by current measurement
if, for all $\O_a$ on $\Si$,
\be\label{10}
(i) \qquad \langle Q\rangle:=\langle\Psi|\hat{Q}[\O]|\Psi\rangle=Q[\O]+
O({\d\over L}).
\ee
\be\label{11}
(ii) \qquad \D_Q:=(\langle Q^2\rangle -\langle Q\rangle^2)^{1\over 2} << Q[\O].
\ee
where $\d$ is a fixed length chosen as $l_{Pl}<\d <<L$. However, this
definition may face obstruction when it is used for non-compact $\Si$, such as
$R^3$. As argued in Ref.\cite{AG99}, the weave states which 
describe the geometries on $R^3$ have to be based on graphs of 
an infinite number of curves, while the states in $\cal H$ constructed so far 
are based on graphs of finite collections of curves. We now think of a way to
overcome the obstruction to a certain extent. 

Suppose there is a cover $\{C_i\}$, consisted of 3-dimensional regions
$C_i$, of a non-compact $\Si$, such that for any $C_i$, a weave state $W_{\Si}$
based on a graph $\G$(may consist of an infinite number of curves) can
always be expressed as: 
\be\label{12}
W_{\Si}=W_{C_i}W_{\Si - C_i},
\ee
where the cylindrical functions
$W_{C_i}$ and $W_{\Si - C_i}$ are based, respectively, on the
subgraphs of $\G$ restricted to $C_i$ and $\Si - C_i$, and the subgraphs of the
regions $C_i$ all consist of finite numbers of curves. Then we can
define that $W_{\Si}$ approximates a classical metric on $\Si$ if all
$W_{C_i}$ approximate, according to Eqs. (\ref{10}) and (\ref{11}), the
metrics restricted to $C_i$. Note that this definition is valid for
all of the weaves states and their 3-metrics appeared so far.

We now construct a weave state which approximate the above given degenerate
metric $h_{ab}$. The basic idea is to consider a family of an infinite number 
of non-intersecting open curves, $\{ \g_i \}$, instead of closed loops on 
$R^3$. All of the $\g_i$ are required to be the
integral curves of the degenerate vector field of $h_{ab}$, and hence match
the $Z$-coordinate curves exactly. This kind of curve was called ``large 
loops'' in Refs.\cite{Ze92,Bo94}. Using the induced 2-metric $q_{ab}$ on a
2-surface $Z=const.$, we fix the intersections of ${\g_i}$ and the surface as
the lattice sites of a square lattice on $R^2$ with lattice spacing $\l$. As
mentioned in Ref.\cite{AG99}, a way of dealing with states based on curves of
infinite length is to consider a compactification of $\Si$ \cite{AG98}. Thus
$\g_i$ may also be regarded as a closed loop on $\bar{R^3}$, where 
$\bar{R^3}:=R^3\cup \ifi$ is the one-point compactification of $R^3$. Follow
Ref.\cite{AG99}, we define the desired "quasi-coherent" state, $W_{\{\}}$, 
based on $\{\g_i\}$ as:
\be\label{13}
W_{\{\}}:=lim_{n\rightarrow\ifi} \prod_{i=1}^n\p_i
\ee
where \be\label{14}
\p_i:=\eta exp\left(\b{ }Tr[\r_1(H[\g_i]-e)]\right),
\ee
here, $\b$ is an arbitrary constant, $e$ is the identity in $SU(2)$, and $\eta$
is a normalisation factor. 

To see if $W_{\{\}}$ weaves the classical geometry
determined by $h_{ab}$, let us consider a cover $\{{\cal O}_m\}$ of $R^3$, 
where ${\cal O}_m$ denotes the 3-dimensional region 
$\{(X,Y,Z)|X^2+Y^2<m^2, m\in N\}$,
here, $N$ is the collection of nature numbers. Let $n$ be the number of curves
$\g_i$ in region ${\cal O}_m$, it is obvious from Eqs. (\ref{13}) and
(\ref{14}) that, for any ${\cal O}_m$,
\be\label{15}
W_{\{\}}=W_nW_{\{\}-n},
\ee
where $W_n$ and $W_{\{\}-n}$ are based, respectively, on the graphs 
$\{ \g_i\sbs{\cal O}_m \}$ and $\{ \g_j\sbs (R^3-{\cal O}_m)\}$, which are 
the subgraphs of $\{ \g_i \}$ restricted, respectively, to ${\cal O}_m$ and 
$(R^3-{\cal O}_m)$, and
\be\label{18}
W_n=\prod_{i=1}^n\p_i.
\ee

The remaining task is to prove that $W_n$ approximates the geometry of 
$h_{ab}$ on
${\cal O}_m$. Calculations similar to that of Ref.\cite{AG99} show that $\p_i$
can be expanded in terms of the eigenstates of $\hat{Q}[\O]$ as:
\be\label{16}
\p_i=\sum_{P=0}^{\ifi}s_P \Psi_P[\g_i],
\ee
where \be\label{17}
s_P={I_P(2\b)-I_{P+2}(2\b)\over \sqrt{I_0(4\b)-I_2(4\b)}},
\ee
here, $I_P(x)$ is the modified Bessel function of order $P$. 

From Eqs. (\ref{7}), (\ref{18}), and (\ref{16}) we obtain the expectation value
of $\hat{Q}[\O]$, 

\begin{eqnarray} \label{19}
\langle W_n|\hat{Q}[\O]|W_n\rangle&=&16\pi l^2_{Pl} \sum_{\g_i,i=1}^n
\sum_{P=0}^{\ifi}s_P^2\sqrt{{P\over 2}({P\over 2}+1)}\int_{\g_i} |\O_Z|dZ
\nonumber\\ 
&=&16 \pi l^2_{Pl}\sum_{P=0}^{\ifi} s_P^2 \sqrt{{P\over 2}({P\over2}+1)}
({1 \over \l^2}) \int_{{\cal O}_m} |\O_Z| dXdYdZ + O({\l \over L}).
\end{eqnarray} 
Let \be\label{19.1}
\l=l_{Pl}\left[ 16\pi\sum_{P=0}^{\ifi}s_P^2\sqrt{{P\over 2}({P\over2}+1)}
\right]^{1\over 2},
\ee
then, from Eqs. (\ref{4}) and (\ref{19}) we have, on region ${\cal O}_m$,
\be\label{20}
\langle W_n|\hat{Q}[\O]|W_n\rangle=Q[\O]+O({\l \over L}).
\ee
Furthermore, straitforward calculations yield
\be\label{21}
[\langle W_n|\hat{Q}^2|W_n\rangle -(\langle W_n|\hat{Q}|W_n\rangle)^2]^
{1\over{2}}=l_{Pl}\xi
\left[ \int_{X^2+Y^2<m^2} dXdY\left( \int |\O_Z|dZ\right)^2 +O({\l\over L})
\right]^{1\over{2}},
\ee
where \be\label{19.2}
\xi=\left[ 16\pi {{\sum_{P=0}^{\ifi}s_P^2 {P\over 2}({P\over 2}+1)-
\left(\sum_{P=0}^{\ifi}s_P^2\sqrt{{P\over 2}({P\over 2}+1)}\right)^2}\over
 {\sum_{P=0}^{\ifi}s_P^2\sqrt{{P\over 2}({P\over 2}+1)}}}\right]^{1\over{2}}.
\ee
Taking account of $\int dXdY(\int |\O_Z|dZ)^2 \sim (\int dXdY\int
|\O_Z|dZ)^2$ and the ordor of $\xi$, Eq.(\ref{21}) is estimated as:
\be\label{22}
[\langle W_n|\hat{Q}^2|W_n\rangle -(\langle W_n|\hat{Q}|W_n\rangle)^2]^
{1\over{2}} \sim l_{Pl}Q[\O]<<Q[\O].
\ee
We conclude from Eqs. (\ref{21}) and (\ref{22}) that $W_{\{\}}$ approximates 
the degenerate metric $h_{ab}$ on $R^3$ at scales larger than $L$.

The concrete values of $\l$ and $\xi$ can be obtained
from Eqs. (\ref{19.1}) and (\ref{19.2}) by fixing a particular value of $\b$.
For example, we have
\be \beta=20: \ \ \lambda= \sqrt{3.545(16\pi l_{Pl}^2)}=13.35l_{Pl}, \ \ \
\xi=\sqrt{{16\pi 1.509\over 3.545}}=4.63; \ee 
\be \beta=40: \ \
\lambda=\sqrt{5.695(16\pi l_{Pl}^2)}=16.92l_{Pl}, \ \ \ \xi=\sqrt{{16\pi
1.164\over 5.695}}=3.21. \ee 

The "quasi-coherent" feature of the weave 
$W_{\{\}}$ can be seen from its construction of Eqs. (\ref{13}), (\ref{14}), 
and (\ref{18}). The functions $W_n$ take on their maximum values when 
$H[\g_i]=e$ and hence, as
$n\rightarrow \ifi$, the function $W_n$ becomes increasingly peaked around the
connections $A_a^i$ which give a trivial holonomy along all curves $\g_i$.

Note that the functions $Q[\O]$ carry sufficient information about the 
3-metric. If
we know $Q[\O]$ for every smooth 1-form $\O_a$, the metric is known completely.
Using the area operator \cite{ARS92,AL97}, it is not difficult to check that
the weave $W_{\{\}}$ will reproduce as well the correct values of the areas of
any 2-surfaces measured by $h_{ab}$ in $R^3$. Since the curves $\g_i$ are 
non-intersecting, $W_{\{\}}$ will give a zero expectation value of the volume
operator\cite{PR96,AL98} for any 3-dimensional regions. This is the right
result because $h_{ab}$ is degenerate.

\section{Evolving a non-degenerate metric from the degenerate one}
\label{sec:3}

We will show in this section that a non-degenerate space-time region
can be evolved by the classical Ashtekar equations from the degenerate
3-metric woven in last section. We would like to use the complex Ashtekar
formalism \cite{As86}, though the real Ashtekar formalism is preferred in
studying quantum states such as that in the last section. Now both of the basic
variables $A_a^i$ and $\Et^a_i$ are complexified. The constraint and evolution
equations take rather simply forms as follows \cite{As86}:
\be\label{23}
{\cal D}_a\Et^a_i=0,\qquad
\Et^a_i F^i_{ab}=0,\qquad
\Et^a_i\Et^b_j F_{abj}\e^{ijk}=0,
\ee
\be\label{24}
\dot{A}_b^i=i\;\hbox{${}_{{}_{{}_{\widetilde{}}}}$\kern-.5em\it N}\Et^a_j
F_{abk}\e^{ijk}+N^a F_{ab}^i,
\ee
\be\label{25}
\dot{\Et}^b_i=-i{\cal D}_a(\;\hbox{${}_{{}_{{}_{\widetilde{}}}}$\kern-.5em\it
N}\Et^{aj} \Et^{bk})\e_{ijk}+2{\cal D}_a(N^{[a}\Et^{b]}_i),
\ee
where ${\cal D}_a$ and $F_{ab}^i$ are, respectively, the derivative operator 
and curvature associated with $A_a^i$, and
$\;\hbox{${}_{{}_{{}_{\widetilde{}}}}$\kern-.5em\it N}$ and $N^a$ are, 
respectively, the lapse density (weight -1) and the sift vector. To recover a
real theory, the reality condition that the metric constructed from $\Et^a_i$
by Eq.(\ref{1}) and its time direvative should be real has to be posed. 

We now construct the desired example by applying some reparametrization
procedure\cite{BJ97,MLK99} to the Minkowski metric.
Consider the Minkowski line element in double-null coordinates $\{ U,V,X,Y\}$:
\be\label{28}
ds^2=-dUdV+dX^2+dY^2.
\ee
Under the reparametrizations $U=U(u)$ and $V=V(v)$, it becomes
\be\label{29}
ds^2=-U'V'dudv+dX^2+dY^2,
\ee
where $U':=dU/du$ and $V':=dV/dv$. In order to get the desired solution, we
define the functions $U$ and $V$ as follows:
\be\label{30}
U(x)=V(x):=
\left\{
\begin{array}{ll}
x^r, & \mbox{ if } x\geq 0 \\
0, &\mbox{ if } x<0
\end{array}
\right.
\ee
where the real number $r\geq 3$. A simple coordinate transformation
\be\label{31}
u=t-Z,\qquad
v=t+Z,
\ee
turns the metric (\ref{29}) into
\be\label{32}
ds^2=U'(t-Z)V'(t+Z)(-dt^2+dZ^2)+dX^2+dY^2.
\ee
Consider this metric on $R^4$ covered by coordinates $\{ t,X,Y,Z\}$, 
Eq.(\ref{30})
means it is non-degenerate in the wedge region $\{ u>0 \}\cap \{ v>0 \}$ and
degenerate outside. The key point is that the space-time with metric
(\ref{32}) represents an evolution of some conjugate pair $(A_a^i, \Et^a_i)$
on $R^3$ covered by $\{ X,Y,Z\}$, satisfying the Ashtekar equations (\ref{23}),
(\ref{24}), and (\ref{25}) as well as the reality condition, and hence is a
solution of the Ashtekar theory. This conjugate pair reads,
\be\label{33}
(A_a^i)=0
\ee
and
\be \label{34}
(\Et^a_i)=\left(
\begin{array}{ccc}
{1\over 2}(U'+V') & {i\over 2}(U'-V') & 0 \\
-{i\over 2}(U'-V') & {1\over2}(U'+V') & 0 \\
0 & 0 & 1
\end{array}
\right),
\ee
where the rows of $(\Et^a_i)$ are $X$, $Y$, $Z$ components, with lapse density
and sift vector
\be\label{35}
\;\hbox{${}_{{}_{{}_{\widetilde{}}}}$\kern-.5em\it N}=1,\qquad
N^a=0.
\ee
It is straightforward to check that Eqs. (\ref{33}), (\ref{34}), and (\ref{35})
indeed satisfy Eqs. (\ref{23}), (\ref{24}), and (\ref{25}), and the space-time
metric constructed from them is the same as Eq.(\ref{32}). The interesting 
feature of this solution is that the 3-metric on $R^3$, being
degenerate initially, becomes partially non-degenerate in the time evolution
as can be seen in the space-time diagram (Figure 1). 

\begin{figure}[htbp]
 \begin{center}
 \input{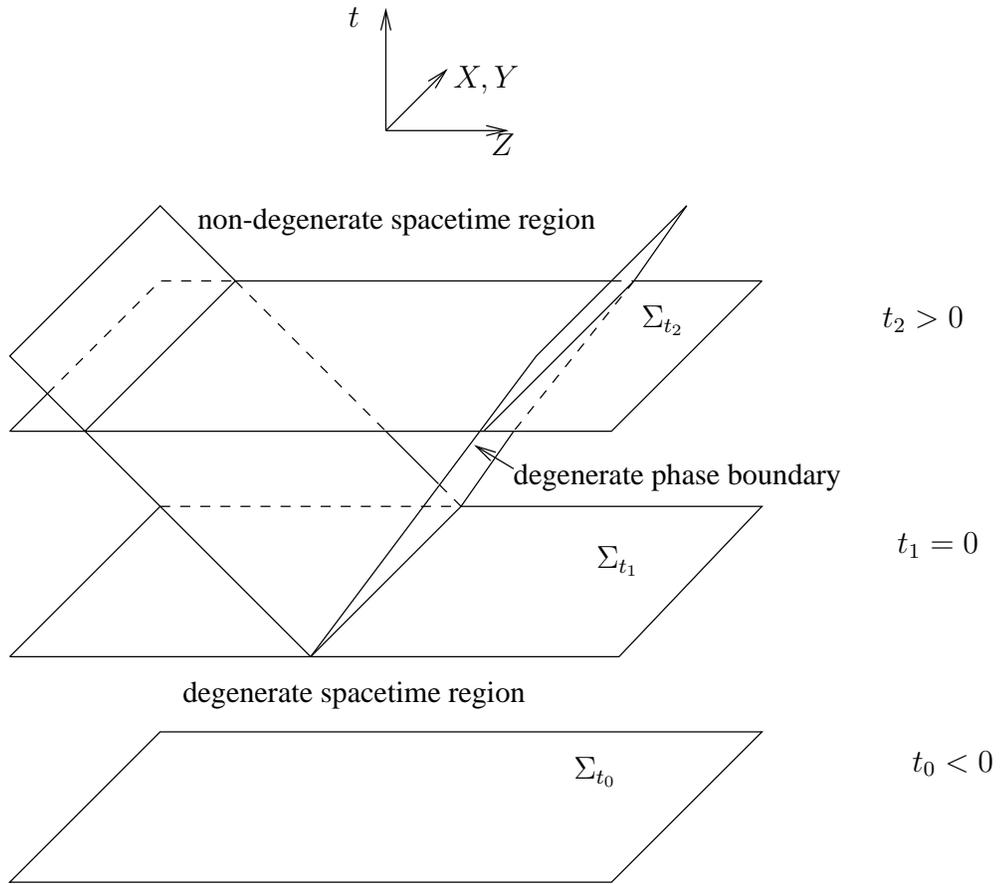}
  \caption{An evolution of a non-degenerate space-time region from the 
degenerate initial data}
 \label{fig:1}
 \end{center}
\end{figure}

The phase boundaries
between the degenerate and non-degenerate regions are null hypersurfaces 
agreeing with the conclusion of Ref.\cite{ML99}. Moreover, before the
non-degenerate metric appears, i.e., for any $t\leq 0$, the induced spatial
metric of Eq.(\ref{32}) is of rank 2 and exactly the same as metric
(\ref{3}) on $R^3$, which is approximated by the weave state $W_{\{\}}$. In
other words, $W_{\{\}}$ has approximated the degenerate 3-metric from which
a non-degenerate metric region can be evolved by the classical Ashtekar
equations. Another approving point is that the $SU(2)$ connection (\ref{33})
gives a trivial holonomy along all curves on $R^3$, and hence is one of the
connections peaked around by the same weave state.
 
Note that metric (\ref{32}) is not $C^{\ifi}$ at surfaces $u=0$ and $v=0$. 
However, one can let the power $r$ in Eq.(\ref{30}) be large enough to obtain 
some desired differentiability. Note also that we could choose other initial 
data of $(A_a^i, \Et^a_i)$ in the gauge 
giving the same $h_{ab}$, from which a degenerate metric on the whole space-
time would be evolved. This supports the observation in Ref.\cite{Ho91} that 
there are gauge transformations which relate degenerate and non-degenerate 
metrics.

\section{Comments and discussion}
\label{sec:4}

The weave state based on ``large loops'' was first proposed in 
Ref.\cite{Ro91} to approximate a flat 3-metric. But, further investigations 
show that this kind of weave can 
not give the ``correct eigenvalue'' for the area operator \cite{Ze92,Bo94}. 
However, our construction shows that ``large loops'' are well suited to weave 
degenerate metrics without any problem for the area operator, because the 
presence of preferred directions of the curves just respects the feature of 
degenerate metrics.

Our example in Section 3 shows that the degeneracy of 3-metrics is not 
preserved by Ashtekar's equations, although it is concluded in 
Ref.\cite{LW979} that the ``degeneracy type of triads'' is locally preserved 
by the 
evolution. Moreover, in contrast to the solution in Ref.\cite{ML98a} where 
the causality has to be broken in order to evolve a degenerate metric from 
non-degenerate initial data, the causal structure, which may be degenerate 
\cite{Ma96}, of the whole space-time can be still well without any breaking 
in the inverse evolution. In this sense the non-degenerate region in the 
example is causally evolved from the degenerate initial data.

It is straitforward to see that the weave state $W_{\{\}}$ solves the quantum 
Hamiltonian constraint. A common point to all different regularisation 
procedures in loop quantum gravity is that the Hamiltonian constraint operator 
acts only on the nodes of spin networks \cite{Th97,Ez96}. From the 
definition (\ref{13}) and Eq.(\ref{16}), it is obvious that $W_{\{\}}$ can be 
expanded by spin network basis as:
\be\label{36}
W_{\{\}}=\sum_{\{ P_i\}} c_{\{ P_i\}}\Psi_{\{ P_i\}},
\ee
where, $\Psi_{\{ P_i\} }=lim_{n\rightarrow \ifi}\prod_{i=1}^n 
\Psi_{P_i}[\g_i]$ is 
based on the spin network $\{ P_i\}$ which is obtained by colouring $P_i$ to 
every $\g_i$, and the sum is over all possible choices of colouring of 
$\g_i$. Since the graph $\{ \g_i\}$ consistes of non-intersecting curves, 
$\Psi_{\{ P_i\} }$ and hence $W_{\{\}}$ are annihilated by the Hamiltonian 
constraint operator. 
In fact, $W_{\{\}}$ can be viewed as a special kind of combinatorial solution 
in Ref.\cite{Ez96}. 

To get the state solving the diffeomorphism constraint, we use the loop 
representation\cite{RS90,RS95b} and define the spin network state 
$\Phi_{K\{ P_i \} }$ on non-intersecting coloured curves $\a_{p'}$ by:
\be \label{37}
\Phi_{K\{ P_i\} }[\a_{P'}]:=\left\{
\begin{array}{ll}
1, & \mbox{ if } \a_{P'}\in K(\{P_i\}) \\
0, & \mbox{ otherwise}
\end{array}
\right.
\ee
where the s-knot $K(\{ P_i \} )$ is the equivalence class of the embedded 
spin 
networks $\{ P_i \}$ under the action of the diffeomorphism group, 
$Diff(R^3)$, on $R^3$, i.e., $\{ P_i \},\{ P'_j \} \in K$, 
if there exists a $\phi\in Diff(R^3)$, such that $\{P'_j\}=\phi \cdot\{P_i\}$.
 Replacing the spin network basis $\Psi_{\{ P_i \} }$ in Eq.(\ref{36}) by the 
diffeomorphism-invariant knot states $\Phi_{K\{ P_i \} }$, we obtain the 
corresponding quantum state:
\be\label{38}
{\cal W}_{K\{\}}=\sum_{\{ P_i \} } c_{\{ P_i \} }\Phi_{K\{ P_i \} },
\ee
which solves all the quantum constraints. Hence, ${\cal W}_{K\{\}}$ should be 
a physical state of loop quantum gravity for $R^3$, although it does not 
belong to the Hilbert space constructed currently for the states based on 
graphs of a finite number of curves. Since 
the spin network $\{ P_i \}$ corresponds to a rank 2 degenerate flat metric 
$h_{ab}$, the equivalence class $K(\{ P_i \} )$ of spin networks should 
correspond 
to the equivalence class of all metrics related to $h_{ab}$ by a spatial 
diffeomorphism. Thus it is natural to interpret ${\cal W}_{K\{\}}$ as 
representing the rank 2 degenerate flat 3-geometry at large scales.

Moreover, the result in Section 3 shows that this degenerate 3-geometry can be 
related to some locally non-degenerate geometry by classical Ashtekar's 
equations, and hence it plays the role of a bridge between a physical state in 
canonical quantum gravity and the familiar classical geometry.

\subsection*{Acknowledgements}
Y. Ma would like to thank Professor O. M. Moreschi and professor
O. A. Reula for helpfull discussions, and acknowledge support from FONCYT BID 
802/OC-AR PICT: 00223. Y. Ling was 
supported by the NSF through grant PHY95-14240, a gift from the 
Jesse Phillips Foundation, and by the 
Department of Physics at Pennsylvania State University.

\end{document}

%%% Local Variables: 
%%% mode: latex
%%% TeX-master: t
%%% End: 